\documentclass[aps,prd,tightenlines,showpacs,amsmath,amssymb]{revtex4}%
\usepackage{graphicx}% Include figure files
\usepackage{epsfig}
\usepackage{graphics,color}
\usepackage{amssymb}
\usepackage{amsmath}
\usepackage{latexsym}
\usepackage{hyperref}
\usepackage{amsbsy}
\usepackage{bm}
\usepackage{url}

\begin{document}
\title{Nucleon Properties in the Polyakov Quark Meson Model}
\author{Yingying Li$^1$}
\author{Jinniu Hu$^2$}
\author{Hong Mao$^1$}
\email {mao@hznu.edu.cn (corresponding author)}
\address{1. Department of Physics, Hangzhou Normal University, Hangzhou 310036, China}
\address{2. School of Physics, Nankai University, Tianjin 300071, China}

%\date{}

\begin{abstract}
We study the nucleon as a nontopological soliton in a quark medium as well as in a nucleon medium in terms of the Polyakov quark meson (PQM) model with two flavors at finite temperature and density. The constituent quark masses evolving with the temperature at various baryon chemical potentials are calculated and the equations of motion are solved according to the proper boundary conditions. The PQM model predicts an increasing size of the nucleon and a reduction of the nucleon mass in both hot environment. However, the phase structure is different from each other in quark and nucleon mediums. There is a crossover in the low-density region and a first-order phase transition in the high-density region in quark medium, whereas there exists a crossover characterized by the overlap of the nucleons in nucleon medium.

\end{abstract}


\maketitle

\section{Introduction}
QCD as a theory of strong interaction is applied to understand how the conversion from hadrons to quark-gluon plasma (QGP) is related to a restoration of the chiral symmetry and a deconfinement. This is a topic of great interest related to the physics of heavy ion collisions (HIC) at ultrarelativistic energies as well as to the processes in the interior of compact stars~\cite{Yagi:2005yb,Fukushima:2010bq,Braun-Munzinger:2015hba}. However, because of the property of confinement at low energy scale,  even though remarkable achievements have been made currently in lattice QCD~\cite{Fodor:2009ax,Ding:2015ona}, the analytical and numerical calculations directly from QCD are hindered. We are still lack of capabilities to describe the low-energy nonperturbative phenomena in the framework of QCD theory, especially when baryons are involved in hadron phase. Therefore, it is usually to apply effective models to study the nonperturbative structure of the QCD vacuum, such as the Nambu-Jona-Lasinio (NJL) model~\cite{Nambu:1961tp,Vogl:1991qt}, the linear sigma model with quarks (or the quark-meson model)~\cite{GellMann:1960np,Scavenius:2000qd}, the Brueckner-Hartree-Fock (BHF) theory\cite{Ring:1980}, and the relativistic mean-field (RMF) model~\cite{Serot:1986,Serot:1997xg}.

The quark-meson (QM) model is the simplest purely quark model which incorporates the chiral symmetry and allows for its spontaneous breaking. At finite temperature and density, based on a nontopological soliton model, the QM model shows a significant success in description of meson and nucleon properties both in vacuum and in a thermal medium~\cite{Birse:1983gm,Kahana:1984dx,Birse:1984js,AbuShady:2012zza,Mao:2013qu,Zhang:2015vva,Mansour:2015yha}. Whereas the bulk thermodynamics and QCD phase structure obtained in the QM model with the mean-field approximation deviate from the lattice QCD simulations and the experimental data. Furthermore, the nontopological soliton model based on the QM model seriously predicted that there is only a first-order phase transition in whole phase plane and the critical temperature is relatively small around $T_c\sim 110$ MeV. In order to cure these problems, the confinement effect associated with the Polyakov loop dynamics was implemented in the QM model~\cite{Schaefer:2007pw,Mao:2009aq,Skokov:2010sf,Gupta:2011ez,Schaefer:2009ui,Fukushima:2017csk}, and the nontopological soliton solutions in the Polyakov quark-meson (PQM) model including the renormalized fermionic vacuum were solved in Ref.~\cite{Jin:2015goa}. In these works, it was shown that the inclusion of the Polyakov loop is necessary and important comparing with the lattice QCD simulations. Quite encouragingly, after assuming that the thermal medium in hadron phase can be approximately replaced by a uniform quark medium, the PQM model really gives a reasonable critical deconfinement temperature about $177$ MeV for zero baryon chemical potential and a standard QCD phase diagram in agreement with the lattice data and other phenomenological models' predictions~\cite{Fukushima:2010bq,Braun-Munzinger:2015hba}.

In this work, we continue to study a nontopological soliton solution of the PQM model at finite temperature and density but with a hot matter considered as a nucleon medium rather than a quark medium adopted in our previous studies in Refs.~\cite{Mao:2013qu,Jin:2015goa}. In this case, we will merely take into account the nuclear degrees of freedom in hadron phase before the QCD phase transition.  The active quark degrees of freedom will be excluded at the nuclear matter density. On the other hand, in hadron phase, at first, the quarks get constituent masses due to the spontaneous chiral symmetry breaking and they have to bind together to form a nucleon as the bound state or soliton due to the confinement. Then, similar to the approach of the constituent quark model~\cite{Isgur:1978xj,Toki:1998hf,Shen:1999um}, the meson fields which mimic nucleon interactions, act on quarks inside a nucleon and change the nucleon properties in nuclear medium. In particular, the $\sigma$ meson is considered as the amount related with the change of the chiral condensate in nuclear medium. Therefore, it is natural to get the reduction of the constituent quark mass inside of the nucleon in a nucleon medium. 

Besides our current works, based on an alternative topological soliton model of the nucleon\cite{Christov:1995vm, Alkofer:1994ph}, the modifications of baryon properties due to the restoration of the chiral symmetry in an external hot and dense thermal medium have been previously investigated in a chiral soliton model with or without explicit vector mesons\cite{Weigel:2008zz}. In these studies, the nucleon now arises as a topological soliton of the NJL model or the Skyrme model\cite{Zahed:1986qz,Meissner:1989bd}, but the parameters of which are chosen to be the medium-modified meson values evaluated within the chiral perturbation theory or the NJL model \cite{Bernard:1989pn,Bernard:1989pj}, respectively. Actually, these two different parameters settings are corresponding to two different ways for the nucleon embedded in a hot thermal medium. In the former case, the nucleon is treated as a baryon-number-one topological soliton immersed in a medium of hot pion gas\cite{Walliser:1997uk}, while in the latter case, the nucleon is to be dipped in a medium of hot quark matter\cite{Berger:1996hc,Schleif:1997pi}. Our present study will extend these previous works to a more realistic scenario by considering the thermal medium in hadron phase as a hot nucleon matter. In such a hybrid nontopological soliton model, similar to the approach of the RMF model, the nucleons as a $B=1$ soliton in a hot medium are coupled directly to the meson fields in order to respect to some extent confinement before the phase transition. This kind of extension will make the present model more suitable for a self-consistent thermodynamical description of the hadron-quark phase transition.

The paper is organized as follows. First we introduce the two flavors Polyakov-loop extended quark meson model and fix its parameters in Sec. II. In Sec. III, we study constituent quark masses at finite temperature and density in present model when a hot medium is considered as the quark and the nucleon medium. The types of the QCD phase transition are also discussed in this section. In Sec. IV, after obtaining the constituent quark mass for the valence quark, we solve the nontopological soliton as a nucleon in quark and nucleon mediums, respectively. Nucleon properties, such as the mass and radius, are also investigated carefully and extensively. The QCD phase structure is also addressed in the end of this section. We conclude with a summary and discussions in the last section.

\section{Model Formulation}
We use the QM model with $N_f=2$ flavor quarks coupled to a spatially constant time-dependent gauge field as background, representing Polyakov loop dynamics to formulate the PQM~\cite{Schaefer:2007pw}. The associated Lagrangian is given as,
\begin{equation}
{\cal L}=\overline{\psi} \left[ i\gamma ^{\mu} D _{\mu}-
g(\sigma +i\gamma _{5}\vec{\tau} \cdot \vec{\pi} )\right] \psi
+ \frac{1}{2} \left(\partial _{\mu}\sigma \partial ^{\mu}\sigma +
\partial _{\mu}\vec{\pi} \cdot \partial ^{\mu}\vec{\pi}\right)
-U(\sigma ,\vec{\pi})-\mathbf{\mathcal{U}}(\Phi,\Phi^*,T) .\\
\label{Lagrangian}
\end{equation}
Here, the chiral part of the Lagrangian with quarks and mesons has $SU_L(2)\otimes SU_R(2)$ symmetry, which is spontaneously broken in the vacuum. $\mathbf{\mathcal{U}}(\Phi,\Phi^*,T)$ represents the temperature dependent effective potential, and it is constructed to reproduce the thermodynamical behavior of the Polyakov loop for the pure gauge case in reasonable agreement with the recent lattice QCD results, and it has the $Z(3)$ center symmetry like the pure gauge QCD Lagrangian.

A possible form of the Polyakov loop potential is the polynomial parametrization based on a Ginzburg-Landau ansatz~\cite{Schaefer:2007pw,Ratti:2005jh},
\begin{eqnarray}
\frac{\mathbf{\mathcal{U}}(\Phi,\Phi^*,T)}{T^4}=-\frac{b_2(T)}{4}(|\Phi|^2+|\Phi^*|^2)-\frac{b_3
}{6}(\Phi^3+\Phi^{*3})+\frac{b_4}{16}(|\Phi|^2+|\Phi^*|^2)^2,
\end{eqnarray}
with the temperature-dependent coefficient $b_2$ defined as
\begin{eqnarray}
b_2(T)=a_0+a_1\left(\frac{T_0}{T}\right)+a_2\left(\frac{T_0}{T}\right)^2+a_3\left(\frac{T_0}{T}\right)^3.
\end{eqnarray}
The parameters in this formula are adjusted to the lattice data for the pure gauge theory thermodynamics, and they are listed as
\begin{eqnarray}
a_0=6.75,\qquad a_1=-1.95,\qquad a_2=2.625, \nonumber \\
 a_3=-7.44,\qquad
b_3=0.75,\qquad b_4=7.5.
\end{eqnarray}
The remaining parameter  $T_0$ for deconfinement in the pure gauge sector is fixed at $T_0=208$ MeV for two flavors\cite{Schaefer:2007pw,Gupta:2011ez}, in agreement with the lattice data. Another possible logarithmic parametrization for the effective potential of the Polyakov loop is also provided by the work in Ref.~\cite{Roessner:2006xn}, however, the particular choice made for this work dose not influence the main conclusions of our work.

Following the standard procedure in the mean-field approximation as given in the work of Ref.~\cite{Scavenius:2000qd}, one can obtain the expression of grand canonical potential as the summation contributions of pure gauge field, meson, and quark/antiquark evaluated in the Polyakov loop,
\begin{eqnarray}\label{omegamf}
\Omega_{\mathrm{MF}}(T,\mu,\sigma,\Phi,\Phi^*) =U(\sigma ,\vec{\pi} )+\mathbf{\mathcal{U}}(\Phi,\Phi^*,T)+\Omega_{\bar{\psi} \psi} ^{\mathrm{reg}}+\Omega_{\bar{\psi} \psi}^{\mathrm{th}} .
\end{eqnarray}
The pure mesonic potential including the $\sigma$ and $\vec{\pi}$ is defined as
\begin{equation}
U(\sigma ,\vec{\pi})=\frac{\lambda}{4} \left(\sigma ^{2}+\vec{\pi} ^{2}
-{\vartheta}^{2}\right)^{2}-H\sigma -\frac{m^{4}_{\pi}}{4 \lambda}+f^{2}_{\pi}m^{2}_{\pi}.
\label{mpot}
\end{equation}
Here $\lambda$ is quartic coupling of the mesonic fields, $\vartheta$ is the vacuum expectation value of scalar field when chiral symmetry is explicitly broken, and the constant $H$ is fixed by the PCAC relation which gives $H=f_{\pi}m_{\pi}^{2}$. The third term in Eq.~(\ref{omegamf}) denotes the renormalized contribution of the fermion vacuum loop, which reads as\cite{Skokov:2010sf}\cite{Gupta:2011ez},
\begin{eqnarray}\label{omegareg}
\Omega_{\bar{\psi} \psi} ^{\mathrm{reg}}=-\frac{N_c N_f}{8\pi^2} M_q^4 \mathrm{ln}(\frac{M_q}{\Lambda}),
\end{eqnarray}
where, $N_f=2$, $N_c=3$ and the modified quark dispersion is $E_q=\sqrt{\vec{p}^2+M_q^2}$. The constituent quark (antiquark) mass $M_q$ is defined as $M_q=g \sigma_v$ together with $\sigma_v\equiv\sqrt{\sigma^2+\vec{\pi}^2 }$. Furthermore, the quark/antiquark contribution in the presence of the Polyakov loop dynamics is written as
\begin{eqnarray}\label{omegapsi}
\Omega_{\bar{\psi} \psi}^{\mathrm{th}} =-2 N_f T \int \frac{d^3\vec{p}}{(2
\pi)^3} \left[ \mathrm{ln} g_q^+ + \mathrm{ln} g_q^- \right].
\end{eqnarray}
The expressions $g_q^+$ and $g_q^-$ are defined as the trace over color space,
\begin{eqnarray}
 g_q^+ &=& \left[ 1+3(\Phi+\Phi^*e^{-(E_q-\mu)/T})\times e^{-(E_q-\mu)/T}+e^{-3(E_q-\mu)/T} \right],  \nonumber\\
 g_q^- &=& \left[ 1+3(\Phi^*+\Phi e^{-(E_q+\mu)/T})\times e^{-(E_q+\mu)/T}+e^{-3(E_q+\mu)/T} \right].
\end{eqnarray}
Here, $\mu$ denotes the quark chemical potentials and it is one third of the baryon chemical potential, $\mu=\mu_B/3$.

In the PQM model, one can get the chiral condensate $\sigma$, and the Polyakov loop expectation values $\Phi$. $\Phi^*$ is a function of $T$ and $\mu$ by searching the global minima of the grand canonical potential, i. e., the derivative thermodynamical potential in Eq.~(\ref{omegamf}) with respect to $\sigma$, $\Phi$, and $\Phi^*$,
\begin{eqnarray}\label{gapeqs}
\frac{\partial \Omega_{\mathrm{MF}}}{\partial \sigma}=0, \qquad \frac{\partial \Omega_{\mathrm{MF}}}{\partial
\Phi}=0, \qquad \frac{\partial \Omega_{\mathrm{MF}}}{\partial \Phi^*}=0.
\end{eqnarray}

We will take the values $f_{\pi}=93$ MeV, corresponding to the pion decay constant and $m_{\pi}=138$ MeV is the pion mass in our numerical computation. Unlike the pion, the mass of the $\sigma$ meson still has a poorly known value, but the most recent result of the Particle Data Group considers that $m_{\sigma}$ can vary from $400$ MeV to $550$ MeV with full width $400-700$ MeV\cite{Patrignani:2016xqp}. The coupling constant $g$ is usually fixed by the constituent quark mass in vacuum within the range of $300\sim 500$ MeV, which gives $g \simeq 3.3\sim 5.3$. To confront with the static properties of nucleon in vacuum, we take $m_{\sigma}=472$ MeV and $g=4.5$ as the typical value. It has been proved in Refs.~\cite{Mao:2013qu, Jin:2015goa} that this parameter set can describe the properties of nucleon in vacuum successfully.

\section {Constituent quark mass in Quark and Nucleon Medium}
We will study the constituent quark masses at finite temperature and density based on two different scenarios. Firstly, the quarks inside the nucleon are embedded in a homogeneous background thermal medium filled with unbound quarks having a constituent mass $M_q$. Secondly, the quarks of the nucleon are immersed in a medium of nucleons owing to confinement, which should be relevant in hadron phase before the QCD phase transition.

These two treatments are corresponding to different roles played by the meson fields in thermal medium. In the former scenario, the meson fields are coupled directly to the constituent quarks, and the chiral condensate leads to a modification of the meson properties. Hence, we are able to directly calculate the meson masses and the constituent quark mass for a given chiral condensate in vacuum as well as in a thermal quark medium. However, for the latter scenario, similarly to the quark-mean field (QMF) model for nuclear matter~\cite{Shen:1999um}, in the hadron level, the mesons are coupled to the nucleons of the Fermi sea. The nucleon effective mass and the mean $\bar{\sigma}$ field are determined by the scalar number densities of the nucleons~\cite{Serot:1984ey,Serot:1997xg}, but in the quark level, the quarks get their constituent masses due to spontaneous chiral symmetry breaking, and the $\bar{\sigma}$ mean field is interpreted as the amount related with the change of the chiral condensate in nuclear medium.

\subsection{Quark medium}
In this subsection, we consider a thermal medium as a Fermi sea of quarks. In this case at finite temperature and density, the single quarks from the Dirac sea are allowed to be excited and occupy levels in the positive part of the Fermi sea but leaving antiquarks in the Dirac sea. After obtaining the grand canonical potential $\Omega_{\mathrm{MF}}(T,\mu,\sigma,\Phi,\Phi^*)$ in the presence of the renormalized fermionic vacuum term, one can easily get the chiral condensate $\sigma$, and the Polyakov loop expectation values $\Phi$ and $\Phi^*$ as a function of the temperature for zero and nonzero quark chemical potentials by solving the couple gap equations in Eq. (\ref{gapeqs}).

\begin{figure}
\includegraphics[scale=0.36]{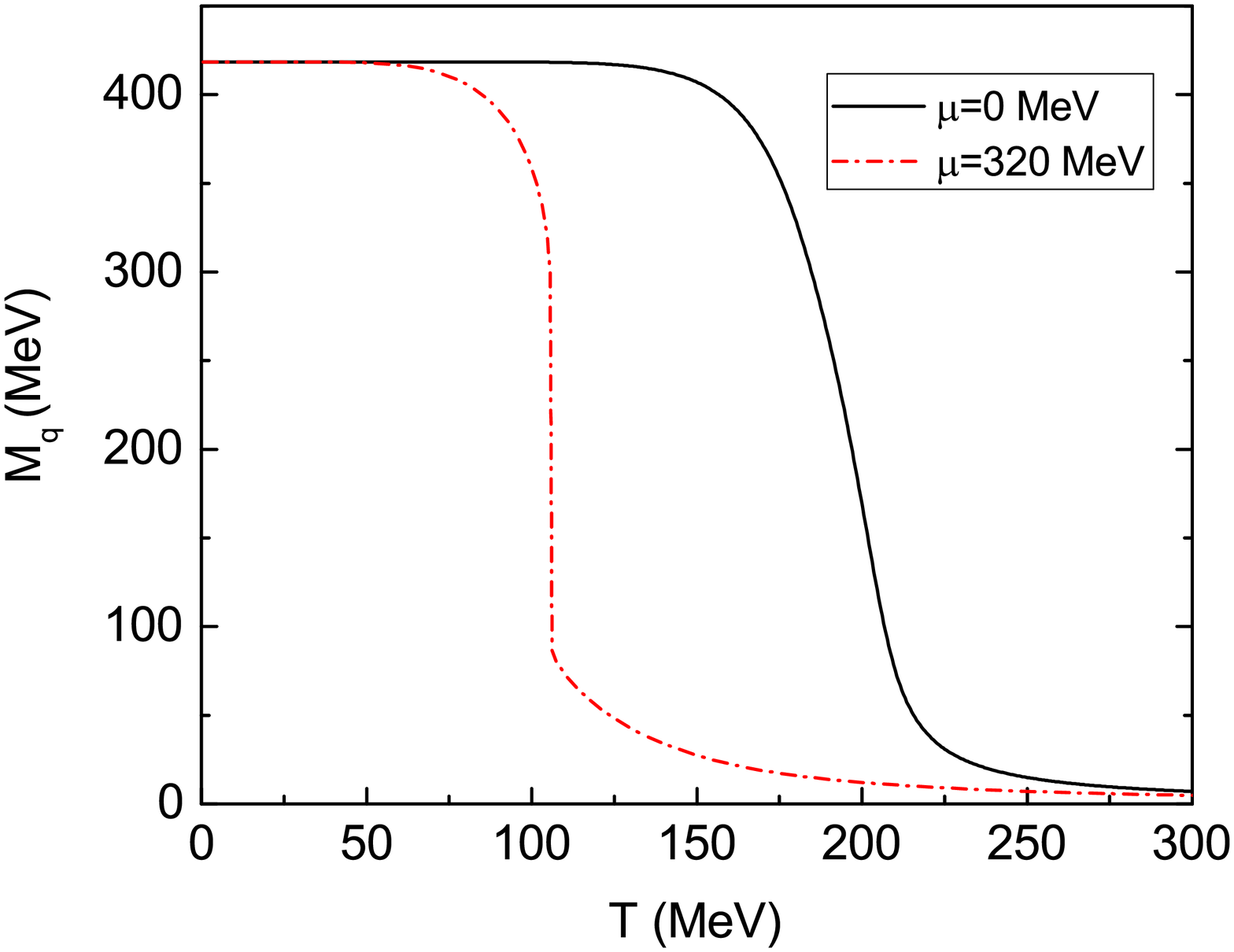}
\caption{\label{Fig01} Constituent quark masses $M_q$ as functions of
temperature for $\mu = 0$ MeV and $\mu = 320$ MeV. The solid curves are for  $\mu = 0$ MeV and the dash-dotted curves are for $\mu = 320$ MeV.}
\end{figure}

The results for the constituent mass $M_q$ as a function of temperature for different quark chemical potentials  are presented in Fig.~\ref{Fig01}. The temperature behaviors of  the constituent masses show that the system experiences a smooth crossover at zero quark chemical potential, while there is a first-order phase transition for some larger quark chemical potential due to the fact that the constituent mass makes jump across the gap of the mass near the critical temperature. Usually, the temperature derivative of  the constituent mass for $u$ quark or $d$ quark has a peak at a specific temperature, which is established as the critical temperature for the chiral phase transition. Hence, for zero quark chemical potential, the chiral restoration occurs at $T_{\chi}^c \simeq 201$ MeV, whereas for a relatively larger quark chemical potential $\mu = 320$ MeV, the critical temperature moves to the lower region around $T_{\chi}^c \simeq 106$ MeV.

For QCD phase diagram, thus, at low quark chemical potential region, it is a crossover between hadronic phase and quark gluon plasma phase, but at  large quark chemical potential region, the PQM model predicts that the phase transition is of first-order and there should exist a so-called QCD Critical End Point (CEP) as the end point of the first-order phase boundary~\cite{Schaefer:2007pw, Jin:2015goa,Gupta:2011ez}.  Finding the signature of the QCD phase transition and the CEP  at high baryon density region are the main goals of the Beam Energy Scan (BES) programs at Relativistic Heavy-Ion Collider (RHIC)~\cite{Aggarwal:2010cw} and the Super-Proton Synchrotron (SPS) facilities~\cite{Abgrall:2014xwa}. We believe that the experimental confirmation of the existence of the CEP will be an excellent verification of QCD theory in the non-perturbative region and a milestone of exploring the QCD phase structure~\cite{Yagi:2005yb, Fukushima:2010bq, Luo:2017faz}.

\subsection{Nucleon medium}
In the previous subsection, we have considered a thermal medium as a quark medium with the constituent quark embedded in. This consideration is acceptable especially for the quark phase after the QCD phase transition in the high energy region where the quarks are set to be free and the single quarks from the Dirac sea are allowed to be excited and occupy levels in the positive part of the spectrum leaving antiquarks in the Dirac sea. However, because of confinement, only the nucleons and mesons are taken as the active degrees of freedom in hadron phase. Thus, as an alternative to the previous consideration, in this subsection we would rather treat the thermal medium as a Fermi sea of nucleons. In this case, the quarks in a nucleon are directly coupled to the scalar and vector meson fields. The nucleon properties change according to the strengths of the mean fields acting on the quarks. In other words, the quarks get constituent masses due to spontaneous chiral symmetry breaking in vacuum and nearly zero mass pions are taken as the Nambu-Goldstone bosons, but in nuclear medium these constituent quark masses are changed accordingly to the mean $\bar{\sigma}$ field, which simulates the interactions between nucleons in dense medium. Therefore, it is natural to get the reduction of the constituent quark mass in the nucleon inside of nuclei or nuclear matter.

To realize above phenomenological viewpoint, we begin with the PQM model of the quark many-body system. In vacuum, the Lagrangian in Eq.~(\ref{Lagrangian}) is reduced to a simple QM model as the Polyakov-loop variables setting to zero, and the thermodynamic grand potential in Eq.~(\ref{omegamf}) becomes a purely mesonic potential equation (\ref{mpot}). When the chiral symmetry is spontaneously broken in the vacuum, the expectation values of the meson fields are $\langle \sigma \rangle=f_{\pi}$ and  $\langle \pi \rangle=0$,  and the constituent quark mass $M_q^{0}$ in vacuum is defined to be
\begin{eqnarray}
M_{q}^{0}=g f_{\pi}=418.5 \mathrm{MeV}.
\end{eqnarray}

After obtaining the constituent quark mass in vacuum, the next step is to generate the nucleon system under the influence of the meson mean fields in nuclear medium. Following the constituent quark model~\cite{Isgur:1978xj} or the QMF model~\cite{Toki:1998hf,Shen:1999um}, the quarks in a nucleon are directly coupled with the scalar $\bar{\sigma}$ mean field, which mimics the attractive parts of the nuclear interaction between nucleons in nuclear matter or nuclei. Assuming the meson mean fields are constant within the small nucleon volume, the constituent quark mass influenced by the $\bar{\sigma}$ mean field  is modified as
\begin{eqnarray}\label{cqmnucleon}
M^*_q=g \sigma_v=M_q^{0}+g\bar{\sigma},
\end{eqnarray}
where the chiral condensate in a nucleon medium is $\sigma_v=f_{\pi}+\bar{\sigma}$.

Unfortunately,  the scalar $\bar{\sigma}$ mean fields can not be self-consistently determined for a given temperature and density in the present PQM model, since the scalar $\sigma$ field and its vacuum expectation value can not serve as the chiral partner of the pion, the generator of the nucleon mass, and the mediator of scalar attraction for nucleons, simultaneously, as discussed in Refs.~\cite{Zschiesche:2006zj,Sasaki:2010bp}. To avoid this problem, in this work we take a simplified approximation by requirement that the chiral condensate is a function of the scalar $\bar{\sigma}$ mean field in Eq.~(\ref{cqmnucleon}) for a nucleon in nuclear medium, this kind of the scenario is also adopted in the QMF model\cite{Toki:1998hf,Shen:1999um}.

To perform the calculations of the scalar $\bar{\sigma}$ mean fields in nuclear matter, we use the RMF model based on local renormalizable Lagrangian densities containing nucleons, neutral scalar ($\bar{\sigma}$) and vector $\omega,~\rho$ mesons. The simple Lagrangian density of the RMF model in nuclear matter is described as
\begin{eqnarray}
{\cal L}
&=&
\bar\psi_N\left[ i\gamma_\mu\partial^\mu-(M_N+g_{\sigma N}\bar\sigma)
-g_{\omega N}\gamma^\mu \omega_\mu
-g_{\rho N} \gamma^\mu\vec\tau_N\cdot\vec\rho_\mu
\right]\psi_N\\\nonumber
&&
-\frac{1}{2} m_\sigma^2{\bar\sigma}^2
-\frac{1}{3} g_2{\bar\sigma}^3
-\frac{1}{4} g_3{\bar\sigma}^4\\\nonumber
&&
+\frac{1}{2} m_\omega^2\omega^2
+\frac{1}{4} c_3\omega^4+\frac{1}{2} m_\rho^2\rho^2,
\end{eqnarray}
where the arrows denote the isospin vectors of $\rho$ meson and $\tau_{N}$ the isospin Pauli operator of nucleon.  The coupling constants between mesons and nucleon, and the strengths of self interaction of mesons, $g_{\sigma N},~g_{\omega N}, ~g_{\rho N}, ~g_2, ~g_3$ and $c_3$ are determined by the empirical saturation properties of infinite nuclear matter and experimental data of stable finite nuclei. In this work, the TM1 parameter set is used, which has achieved a lot of successes in description of the nuclear many-body system~\cite{Sugahara:1993wz}.

The equations of motion about nucleons and mesons can be obtained from the Euler-Lagrange equations. However, in these equations of motion, the quantum fields cannot be solved exactly for complicated many-body system. The no-sea approximation and mean-field approximation are adopted to consider the mesons as classical fields in RMF model. Furthermore, in the infinite nuclear matter, the system has the translational invariance.  Finally, the equations of motion of nucleon and mesons are given as,
\begin{eqnarray}
&&\left[\vec\alpha\cdot\vec k+\beta M^*_N
+g_{\omega N}\omega
+g_{\rho N}\rho\tau_{N,3}\gamma^0\
\right]\psi_{Nk}
=\varepsilon_{Nk}\psi_{Nk}
\end{eqnarray}
and
\begin{eqnarray}
&&m_\sigma^2\bar\sigma+g_2\bar\sigma^2+g_3\bar\sigma^3
=-g_{\sigma N}
\langle\bar\psi_N\psi_N\rangle,\\\nonumber
&&m_\omega^2\omega+c_3 \omega^3=
g_{\omega N}\langle\bar\psi_N\gamma^0\psi_N\rangle,\\\nonumber
&&m_\rho^2\rho=
g_{\rho N}\langle\bar\psi_N\tau_{N,3}\gamma^0\psi_N\rangle,
\end{eqnarray}
where $M^*_N$ is the effective nucleon mass related to $\bar\sigma$ field,
\begin{eqnarray}
M^*_N=M_N+g_{\sigma N}\bar\sigma.
\end{eqnarray}
At finite temperature, the scalar density and vector density are expressed, respectively,
\begin{eqnarray}
\rho_s=\langle\bar\psi_N\psi_N\rangle=\gamma \int \frac{d^3\vec k}{(2\pi)^3} \frac{M^*_N}{\sqrt{k^2+M^{*2}_N}}[f^+_N(T,k)+f^-_N(T,k)]
\end{eqnarray}
and
\begin{eqnarray}\label{netdensity}
\rho=\langle\bar\psi_N \gamma^0 \psi_N\rangle=\gamma\int \frac{d^3\vec k}{(2\pi)^3} [f^+_N(T,k)-f^-_N(T,k)],
\end{eqnarray}
where $\gamma$ is the summations of spin and isospin and $\gamma=4$ for symmetric nuclear matter. $f^{\pm}_N(T,k)$ are the Fermi-Dirac distributions to nucleon and antinucleon defined as,
\begin{eqnarray}
f^{\pm}_N(T,k)=\frac{1}{1+\exp[(\sqrt{k^2+M^{*2}_N}+g_{\omega N}\omega+g_{\rho N}\tau_{N,3}\rho\mp\mu_N)/T]}.
\end{eqnarray}
Here, $\mu_N$ are the proton and neutron chemical potentials, respectively. These equations about nucleon and mesons can be solved self-consistently with numerical method.

\begin{figure}
\includegraphics[scale=0.36]{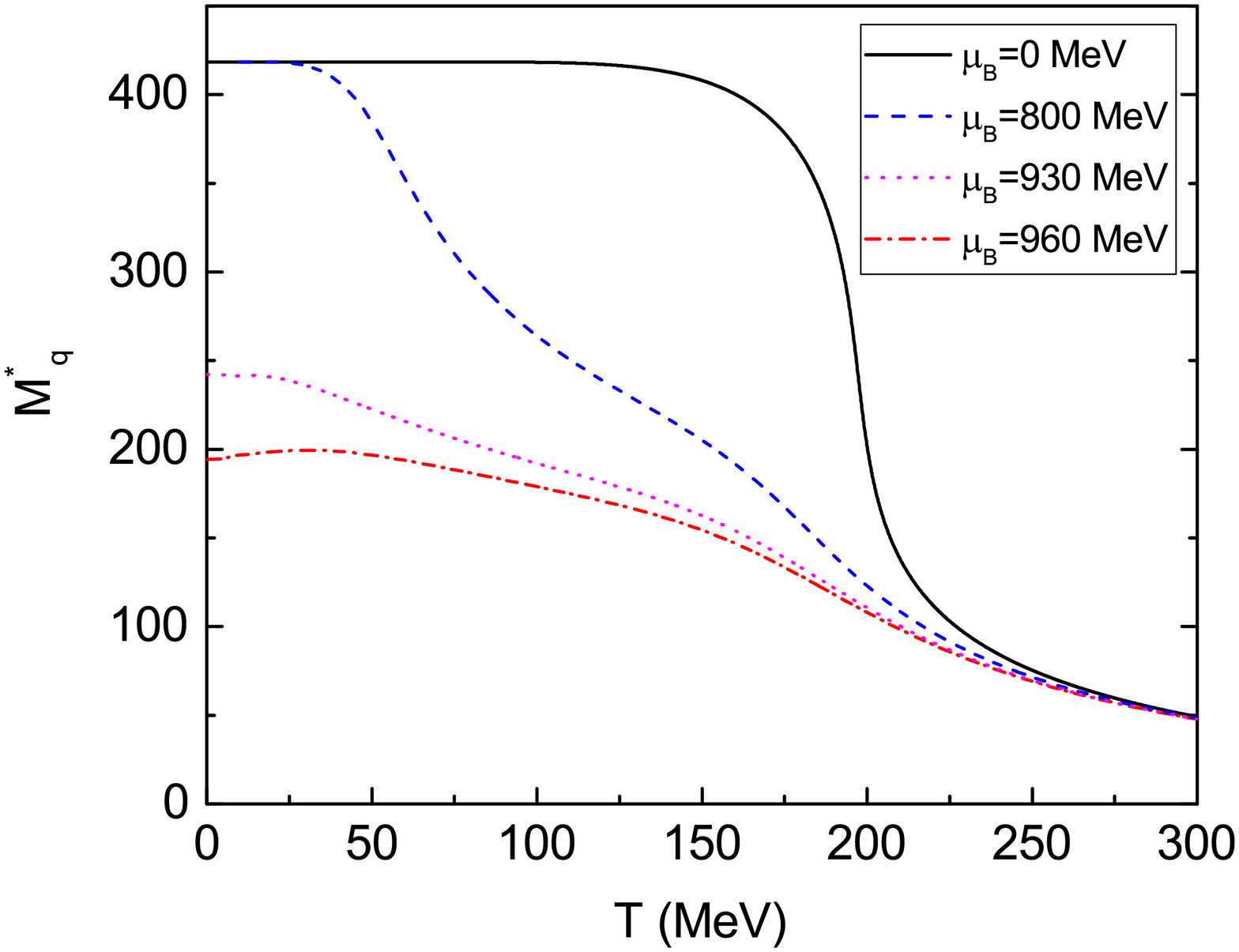}
\caption{\label{Fig02} Constituent quark masses $M^*_q$ in nuclear medium as functions of
temperature for $\mu_B = 0$ MeV, $\mu_B =800$ MeV, $\mu_B = 930$ MeV, and $\mu_B = 960 \mathrm{MeV}$. The solid curves are for  $\mu_B = 0$ MeV, the dashed curves are for $\mu_B =800$ MeV, the dotted curves are for $\mu_B =910$ MeV and the dash-dotted curves are for $\mu_B = 960$ MeV.}
\end{figure}

As long as the scalar $\bar{\sigma}$ mean fields are solved in the RMF model with the TM1 parameter set, the results for $M^*_q$ are presented in Fig.~\ref{Fig02} at finite density as a function of $T$. The behavior of  the constituent quark mass as a function of the temperature characterizes a crossover for both zero and high baryon chemical potentials. As described as in the case of the nucleon in a quark medium in previous subsection, the temperature derivative of the constituent quark mass usually has a peak at some specific temperature, which is established as the critical temperature for the QCD chiral phase transition. Then for zero baryon chemical potential, the QCD phase transition occurs at $T^c=194.8$ MeV, this value is very close to that of quark medium. It means that the use of quark medium instead nucleon medium  is a reasonable approximation for low baryon densities. Whereas, for the high baryon densities, this is not true. For the quarks in quark medium, the corresponding phase transition is rather sharp and the constituent quark mass show a discontinuity at the critical temperature. There is a first-order QCD phase transition for larger quark chemical potential. However, when the quarks are embedded in nucleon medium, the constituent quark mass as a function of the temperature shows a rather smooth behavior for high baryon chemical potentials, therefore, it is impossible to define the critical temperature of a crossover through on the judgment of peaks in the temperature derivative of the constituent quark mass as done in the case of a quark medium. Thus, an alternative way to define the critical temperature for QCD deconfinement phase transition should be addressed carefully and properly.

\section {Nucleon as a Nontopological Soliton at Finite Temperatures and Densities}
We now solve the $B=1$ nontopological soliton as a nucleon in quark and nucleon medium, respectively. The starting points for these two circumstances are the PQM model in vacuum~\cite{Birse:1983gm,Kahana:1984dx}, where the Polyakov loop variables ${\Phi}$, ${\Phi}^*$ are zero and the thermodynamic grand potential $\Omega_{\mathrm{MF}}$ reduces to a purely mesonic potential,
\begin{eqnarray}
\Omega_{\mathrm{M}}(\sigma ,\vec{\pi} )=U(\sigma ,\vec{\pi} )+\Omega_{\bar{\psi} \psi} ^{\mathrm{reg}}.
\end{eqnarray}

In the mean-field approximation, the quarks in a nucleon satisfy the following Dirac equation:
\begin{eqnarray}
-i \vec{\alpha}\cdot \vec{\nabla}\psi(\mathbf{r})-g \beta M_q \psi(\mathbf{r})=\epsilon
\psi(\mathbf{r}), \label{eom1}
\end{eqnarray}
where due to the spontaneous chiral symmetry breaking, the constituent quark mass is defined as
\begin{eqnarray}
	M_{q}=g \sqrt{\sigma(\mathbf{r})^2+\vec{\pi}(\mathbf{r})^2}.
\end{eqnarray}
The ground state of the nontopological soliton is that three quarks occupy in the same lowest Dirac state, with energy $\epsilon$. In order to obtain the solutions of minimum energy, we adopt a hedgehog ansatz, where the meson fields are spherically symmetric and valence quarks are in the lowest $s$-wave level,
\begin{eqnarray}
\sigma = \sigma(r), \vec{\pi} = \hat{\mathbf{r}}\pi(r),\\
\psi = \left(\begin{array}{c} u(r) \\
i \vec{\sigma} \cdot\mathbf{\hat{ r}}v(r)
\end{array}\right)\chi, \label{sltconf}
\end{eqnarray}
where $\chi$ is a state in which the spin and isospin of the quark couple to zero:
\begin{eqnarray}
(\vec{\sigma}+\vec{\tau})\chi=0.
\end{eqnarray}

The mean-field equations of motion for the quark and meson fields are then
\begin{eqnarray}
\frac{du(r)}{dr} =  -\left(\epsilon+g \sigma(r)\right)v(r)-g \pi(r)u(r), \label{equation1}\\
\frac{dv(r)}{dr} =  -\left(\frac{2}{r}-g\pi(r)\right)v(r)+\left(\epsilon-g \sigma(r)\right)u(r), \label{equation2} \\
\frac{d^2 \sigma(r)}{dr^2}+\frac{2}{r}\frac{d\sigma(r)}{dr}-\frac{\partial \Omega_{\mathrm{M}}(\sigma ,\vec{\pi} )}{\partial \sigma}= Ng\left(u^2(r)-v^2(r)\right),\label{equation3}\\
\frac{d^2 \pi(r)}{dr^2}+\frac{2}{r}\frac{d\pi(r)}{dr}-\frac{2 \pi(r)}{r^2}-\frac{\partial \Omega_{\mathrm{M}}(\sigma ,\vec{\pi} ) }{\partial \pi} = -2Ng u(r) v(r).
\label{equation4}
\end{eqnarray}
At $r=0$, $d\sigma/dr$ and $v$ must vanish, while as $r\rightarrow \infty$, $\sigma$ tends to its vacuum value $f_{\pi}$, and $u$, $v$, and $\pi$ tend to zero. Using the asymptotic forms of these equations, the boundary conditions on the fields are given by the following equations from the requirement of finite energy,
\begin{eqnarray}
v(0)=0,     \frac{d\sigma(0)}{dr}=0,\pi(0)=0 \label{boundary1},\\
u(\infty)=0,\sigma(\infty)={\it f}_{\pi},\pi(\infty)=0. \label{boundary2}
\end{eqnarray}
The quark functions should satisfy the normalization condition
\begin{eqnarray}\label{norm}
4\pi \int r^2 \left(u^2(r)+v^2(r)\right)dr=1.
\end{eqnarray}
This set of equations does not has analytic solutions, but is readily solved numerically. Various numerical packages are available for the solution of such equations; one which has been widely used in this field is COLSYS~\cite{Ascher:1981}. The model has two adjustable parameters $g$ and $m_{\sigma}$ which can be chosen to fit various baryon properties, such as the baryon mass $M_N$, the root mean square (RMS) charge radius of proton $r_{cp}$, the magnetic moment $\mu_p$ and the axial to vector coupling constants $g_A/g_V$, which have been measured experimentally. To perform the calculation of all these physical quantities, we use the following definitions,
\begin{eqnarray}\label{energy}
E=M_N=N \epsilon+4\pi\int r^2 \left[ \frac{1}{2}
\left(\frac{d\sigma}{dr}\right)^2+\frac{1}{2}\left(\frac{d\pi}{dr}\right)^2+\frac{\pi^2}{r^2}+\Omega_{\mathrm{M}}(\sigma ,\vec{\pi} ) \right]dr,\\
\langle r_{cp} ^2 \rangle=4\pi\int^\infty_0 r^4(u^2(r)+v^2(r))dr,\\
\mu_p=\frac{8\pi}{3}\int^\infty_0 r^3 u(r)v(r)dr,\\
\frac{g_A}{g_V}=\frac{20\pi}{3}\int^\infty_0
r^2(u^2(r)-\frac{1}{3}v^2(r))dr.
\end{eqnarray}

\subsection{Quark medium}
First we consider the case of a $B=1$ soliton in a thermal quark medium filled with quarks having the constituent mass $M_q$. In this case, the soliton or the nucleon is treated as a localized bound state of three constituent quarks with the interactions of mesons. As long as the unbound constituent quarks are allowed to be inserted into the nucleon, the mean-field values for the $\sigma$ and $\pi$ fields and thereby the constituent quark masses in Eq.~(\ref{omegapsi}) at finite temperature and density could be also obtained by minimizing the relevant thermodynamic ground potential $\Omega_{\mathrm{MF}}$ in Eq.~(\ref{omegamf}) with respect to $\sigma$ and $\vec{\pi}$, therefore, a new set of coupled equations for $\sigma$ and pion fields should be rewritten as,
\begin{eqnarray}
\frac{d^2 \sigma(r)}{dr^2}+\frac{2}{r}\frac{d\sigma(r)}{dr}-\frac{\partial\Omega_{\mathrm{MF}}}{\partial\sigma} = Ng\left(u^2(r)-v^2(r)\right),
\label{equation5}\\
\frac{d^2 \pi(r)}{dr^2}+\frac{2}{r}\frac{d\pi(r)}{dr}-\frac{2 \pi(r)}{r^2}-\frac{\partial\Omega_{\mathrm{MF}}}{\partial\pi} = -2Ng u(r) v(r).
\label{equation6}
\end{eqnarray}
Accordingly, the boundary condition for $\sigma(r)$ in Eq.~(\ref{boundary2}) should be modified as: $r\rightarrow \infty$,
$\sigma(r)$ approaches to the expectation value $\sigma_v$, where thermodynamic grand potential $\Omega_{\mathrm{MF}}$ has an absolute minimum.

In PQM model, because there is only one minimum in the effective Polyakov-loop potential by fixing the chiral order parameters on their expectation values as long as the temperature $T$ is smaller than the critical temperature $T_0$ for deconfinement in the pure gauge sector. Both Polyakov-loop variables $\Phi$ and $\Phi^*$ can not develop a bag-like soliton solutions for $T<T_0$. Hence, the Polyakov loop variables ${\Phi}$, ${\Phi}^*$ will always own their expectation values in whole space, so that they do not increase the equations of motion for the nontopological soliton solutions and should be merely regarded as homogeneous background thermal fields on top of which the chiral soliton is going to embed. Consequently, the properties of a soliton emerged in a thermal quark medium can be studied by solving a set of four coupled Euler-Lagrange equations as well as the case of the PQM in vacuum， where two of these equations are the Dirac equation of the quarks in the Eqs.~(\ref{equation1}) and (\ref{equation2}). The others arise from the thermodynamic grand potential as a set of gap motions for the $\sigma$ meson and pion fields in Eqs.~(\ref{equation5}) and (\ref{equation6}).

In PQM model, the nucleon arises as a nontopological soliton of the bounded constituent quarks, and the soliton at finite temperature and density is a solution of the four Euler-Lagrange equations of motion, Eqs.~(\ref{equation1}), (\ref{equation2}), (\ref{equation5}), and (\ref{equation6}) with the proper normalization condition and the appropriate boundary conditions. This is unlike to the case of the nucleon in vacuum. The situation here will become more complicated when we consider that the soliton immersed in a thermal quark medium. Since the unbound constituent quarks  are treated as the homogeneous background thermal fields with $T$ and $\mu$, which will bring an additional contribution to the total baryon density as long as they are allowed to penetrate into the soliton by the requirement of the equations of motion of the soliton in the Eqs.~(\ref{equation5}) and (\ref{equation6}). Thus, in order to ensure the solitonic baryon number exactly to one, the normalization condition equation (\ref{norm}) should be modified accordingly as
\begin{eqnarray}\label{norm2}
4\pi \int r^2 \left(u^2(r)+v^2(r)\right)dr=1-B_m,
\end{eqnarray}
with
\begin{eqnarray}
B_m=4\pi \int_V \rho_B r^2 dr.
\end{eqnarray}
Here, $\rho_B= -\frac{1}{3}\frac{\partial \Omega_{\mathrm{MF}}}{\partial \mu}$ and $V$ is the volume of the soliton with the RMS charge radius $r_{cp}$.

\begin{figure}
\includegraphics[scale=0.36]{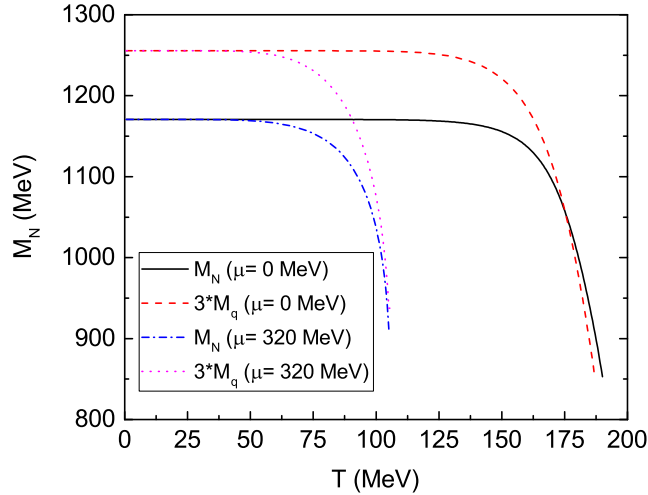}
\caption{\label{Fig03} The total energy of system $M_N$ and the energy of the three free constituent quark $3M_q$ are　given as functions of the temperature $T$. Here one set is for $\mu=0$ MeV, and another set is for $\mu=320$ MeV.}
\end{figure}

As shown in our previous study~\cite{Jin:2015goa}, the effective potential $\Omega_{\mathrm{MF}}$ in Eq.~(\ref{omegamf}) with chiral symmetry breaking phase always supports the existence of the stable soliton solution for the meson fields for both the crossover and the first-order phase transitions. However, the stability of the soliton solution should must be examined carefully by comparing the energy of the soliton (bound state) with that of three constituent quarks in their free states.

In Fig.~\ref{Fig03}, we plot the total energy of system $M_N$ and the energy of three free constituent quarks $3M_q$ at zero and finite chemical potential for different temperatures. These two chemical potentials correspond to the typical crossover and the first-order phase transitions in the QCD phase diagram, respectively. In Fig.~\ref{Fig03}, we can see that the critical temperature for the deconfinement phase transition is lower than that of the chiral phase transition for a crossover, but for the first-order phase transition, the critical temperature for the deconfinement phase transition is coincident with that of the chiral phase transition.

\begin{figure}
\includegraphics[scale=0.36]{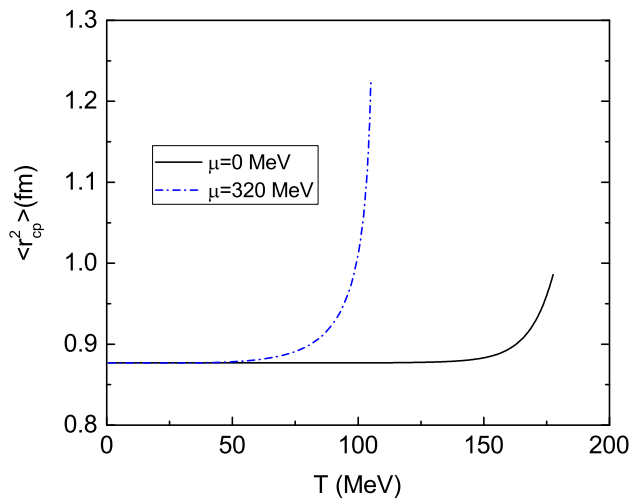}
\caption{\label{Fig04} The proton charge RMS radius of a stable chiral soliton as a function of temperature $T$ at $\mu=0$ MeV and $\mu=320$ MeV. The solid curve is for $\mu=0$ MeV while the dash-dotted curve for $\mu=320$ MeV.}
\end{figure}

At the end of this subsection, the proton charge RMS radius $r_{cp}$ of a stable soliton as a function of temperature for different chemical potentials are plotted in Fig.~\ref{Fig04}. It is shown that the size of the nucleon is going to swell up with the increasing of the temperature slightly at first. When $T$ approaches the critical  temperature for the deconfinement phase transition, $r_{cp}$ grows sharply and then disappears soon due to the delocalization of the soliton in this model.

\subsection{Nucleon medium}
In this subsection, we now consider the soliton in the PQM model with three valence quarks interacting in the nucleon Fermi sea via the meson fields. As in the case of a quark medium, only the $\sigma$ field can develop a non-zero expectation value in the nucleon medium, and the quarks in a nucleon have the constituent quark masses in the form of Eq.~(\ref{cqmnucleon}). Therefore, similarly to the cases in quark thermal medium shown in Eqs.~(\ref{equation5}) and (\ref{equation6}), the corresponding equations of motion for the meson fields  in nucleon medium can be directly rewritten with a finite-temperature part of the relevant thermodynamic ground potential $\Omega_{\mathrm{MF}}$ in Eq.~(\ref{omegamf}) now in terms of nucleons
\begin{eqnarray}
\Omega_{\mathrm{\overline{N}N}}=- \gamma T \int \frac{d^3 \vec k}{(2\pi)^3}  [\mathrm{ln} (1+e^{-(E_N-\mu_B)} ) +\mathrm{ln} (1+e^{-(E_N+\mu_B)} ) ]
\end{eqnarray}
together with $E_N=\sqrt{\vec k^2+{g^2 (\sigma^2+\vec{\pi}^2)}}$ and the net baryon density $\rho_B=\rho=\langle\bar\psi_N \gamma^0 \psi_N\rangle$.

Accordingly, the scalar and pseudoscalar densities of valence quarks and antiquarks representing the thermal medium effects in the equations of motion for meson fields can be formally expressed as,
\begin{eqnarray}
\rho^N_s=\langle \bar{\psi}_N\psi_N\rangle = g \sigma \gamma  \int  \frac{d^3  \vec k}{(2\pi)^3} \frac{1}{E_N} \left[ \frac{1}{ (1+e^{-(E_N-\mu_B)} )} +\frac{1}{ (1+e^{-(E_N+\mu_B)} )} \right ] , \\
\rho^N_{ps}=\langle \bar{\psi}_N i \gamma_5 \vec{\tau}\psi_N\rangle= g \vec{\pi} \gamma  \int  \frac{d^3 \vec k }{(2\pi)^3} \frac{1}{E_N} \left[ \frac{1}{ (1+e^{-(E_N-\mu_B)} )} +\frac{1}{ (1+e^{-(E_N+\mu_B)} )} \right ].
\end{eqnarray}
These densities generate the source terms in the equations of motions for the meson fields. Unlike the case in quark thermal medium, both the scalar and pseudoscalar densities of nucleons and antinucleons in above equations cannot be obtained from the PQM model or the RMF model self-consistently. They have to be considered as input parameters with the constraint that we require $\sigma$ field to asymptotically approach the expectation value $\sigma_v$ in the physical vacuum as given in Eq.~(\ref{cqmnucleon}), while other fields are set to zero.

Under the above treatment, the PQM model now is simplified as the classical chiral soliton model and the well-known results already exhibited in the chiral soliton model in Ref.~\cite{Mao:2013qu} can be duplicated with the exception that the nucleons do not penetrate into the solitons unless the nucleons start to overlap with each other when the temperature is nearby the critical temperature for QCD deconfinement phase transition. Thus the $B=1$ soliton solution could be obtained by solving the Dirac equations together with meson equations of motion which contain the finite-temperature nucleon part, and there will always exist a stable soliton solution for the meson fields in the chiral symmetry-breaking phase. Moreover, such a baryonic phase composed by the solitonic nucleons are stable and they are satisfied with the energy requirement $M_N<3 M_q$ in hadron phase.

\begin{figure}
\includegraphics[scale=0.36]{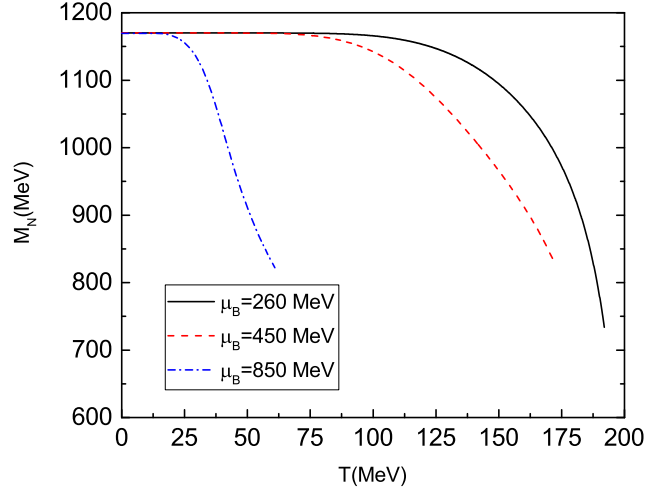}
\caption{\label{Fig05} The total energy of system $M_N$ in nucleon medium as a function of the temperature $T$ for different baryon chemical potentials $\mu_B=260$ MeV, $\mu_B=450$ MeV, and $\mu_B=850$ MeV.}
\end{figure}

The results for the total energy of system $M_N$ in nucleon medium as a function of the temperature $T$ at various baryon chemical potentials are presented in Fig.~\ref{Fig05}. The terminal point for each curve corresponds to the critical temperature when nucleons start to overlap each other, and this critical temperature is set to be a critical temperature for the deconfinement phase transition as described in the following discussions. In Fig.~\ref{Fig03} and Fig.~\ref{Fig05}, at low baryon chemical potential values,  the curves have similar trends for both nucleon and quark medium which means that the use of quark medium instead of nucleon medium is a reasonable approximation. However, this is not true at higher baryon chemical potentials. At large $\mu_B$, both the constituent quark mass and the total energy of system $M_N$ in quark medium show discontinuities at the critical temperature, whereas, those of nucleon medium are quite soft against thermal fluctuations and the corresponding deconfinement phase transition is smooth crossover.

\begin{figure}
\includegraphics[scale=0.36]{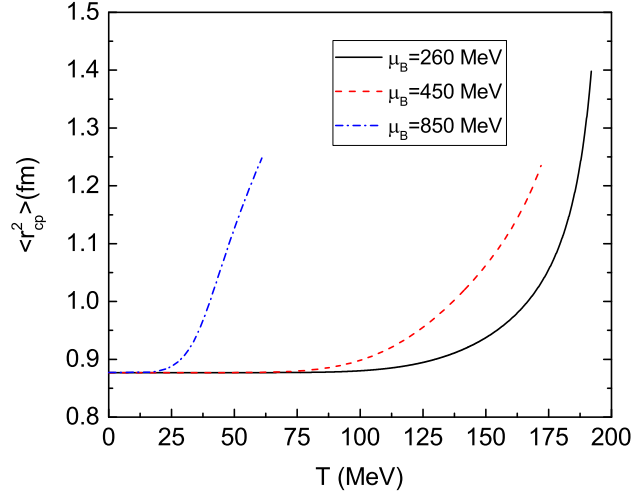}
\caption{\label{Fig06} The proton charge RMS radius of a stable chiral soliton as a function of temperature $T$ for different baryon chemical potentials $\mu_B=260$ MeV, $\mu_B=450$ MeV, and $\mu_B=850$ MeV.}
\end{figure}

In order to illustrate the change of the soliton structure in nucleon medium, we also plot  the proton charge RMS radius $R$ of a stable soliton as a function of temperature for different chemical potentials in Fig.~\ref{Fig06}. All curves in Fig.~\ref{Fig06} show a clear trend  to grow up with the increasing of the temperature. When $T$ is close to the critical temperature, the radius expands quickly which can also be taken as a sign for delocalization of the soliton. In contrast to the case of quark medium in Fig.~\ref{Fig04} , for high baryon chemical potential value, the largest soliton radius is smaller than that of low baryon chemical potential. It means that the nucleons are much easy to get overlapped when $\mu_B$ is large.

After obtaining the radius of the nucleon, we can now investigate the phase diagram of the Polyakov quark-meson model in nucleon medium. Based on above discussions, the nucleon is taken as a classical stable soliton solution of the PQM model in hadron phase,  because the valence quark in a nucleon will also interact with the meson mean field due to the nuclear interaction between nucleons. The constituent quark mass should develop a medium-modified values at finite temperature and baryon chemical potential as shown in Eq.~(\ref{cqmnucleon}). In this prescription, the scalar $\sigma$ meson field simultaneously plays two roles. One is to provide the chiral partner of the pion when dealing with the spontaneous chiral symmetry breaking, and the other is the mediator of scalar attraction for nucleons. Thus, in contrast to the case of the PQM model in quark medium, all valence quarks are going to bind together to form nucleons due to local interactions with meson fields in hadron phase because of confinement. These bound state nucleons are stable against thermal fluctuations contributed from nuclear medium. Since the curves of the constituent quark masses evolving with the temperature are rather smooth, especially for hight baryon chemical potential, we can not use the traditional way used in above to define the phase boundary of hadron phase by setting the derivatives of $M_q$ respect with $T$. Moreover, the studies of masses and radii of the nucleon in above do not remedy the situation. Therefore, we will introduce an alternative way to define the phase boundary of hadron phase in this work.

\begin{figure}
\includegraphics[scale=0.36]{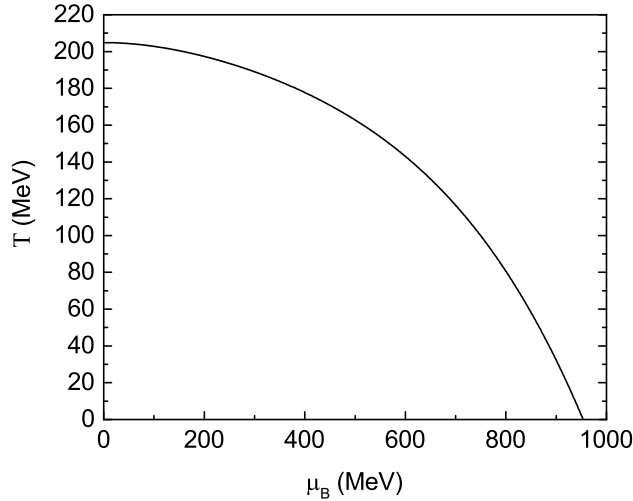}
\caption{\label{Fig07} The phase diagram in the $T-\mu_B$ plane in the Polyakov quark-meson model based on the nontopological soliton picture in nuclear medium. The solid line indicates that nucleons are starting to overlap each other and it gives the phase boundary for hadron phase.}
\end{figure}

It is well known that one of distinguished advantage of the soliton model is to study the overlap of the nucleons. By using this physical picture, in the beginning, the nucleon is well separated from other nucleons, the average net baryon density $\rho_B$ is close to zero in the vacuum and the net baryon density inside the nucleon is almost a constant, which can be roughly defined as
\begin{eqnarray}
\rho_I=\frac{1}{\frac43 \pi R^3},
\end{eqnarray}
here $R$ is the radius of the nucleon/soliton in vacuum. However, when a nucleon is trapped in a nucleon medium, with the increasing of the temperature and density, the net baryon density inside the nucleon starts to decrease accompaniment for the expansion of the nucleon, while the average net baryon density nuclear medium given by RMF model in Eq.~(\ref{netdensity}) is going to deviate at zero and rise very sharply. When the values of two net baryon density cross each other, i.e. $\rho_B \geq \rho_I$, we believe that the nucleon get lots of chances to overlap. This moment is also set as the CEP where hadron phase is changing to quark phase. The corresponding critical temperature is defined as the one for deconfinement phase transition.  The result of the $T-\mu_B$ phase structure from the PQM model in a nucleon medium is plotted in Fig.~\ref{Fig07} . The solid line indicates that nucleons are starting to overlap each other and it gives the phase boundary for hadron phase. Since the deconfinement phase transition in this case is crossover for both low and high baryon chemical potential, we do not find the dramatic structure change for the nucleon when $T$ is close to the critical temperature. By comparison with the results presented in our previous study in Ref.~\cite{Jin:2015goa}, in which phase diagram is constructed based on the picture that the nucleons are trapped in quark medium, for low baryon chemical potential and the phase boundary for the deconfinement phase transition is very close to each other. It means that the dramatic structure change for the nucleon also accompanied with the overlap of the nucleons when $T\simeq T_c$. However, for relatively larger baryon chemical potential, if we treat thermal medium as a system of nucleons, the overlaps are more likely to happen instead of the dramatically fast structure change for nucleons when the temperature is nearby the critical one for the deconfinement phase transition.

\section{Summary and discussion}
We have investigated the nucleon as a $B=1$ soliton in a hot medium using the Polyakov quark-meson model in the mean-field approximation. The constituent quark mass, the mass and radius of the nucleon at finite temperature and density are studied when the hot medium is taken as a quark medium as well as a nucleon medium, separately. Our results show that both the effective nucleon mass and the proton charge RMS radius are not drastically altered according to the increasing of the temperature for a nucleon in a quark medium or in a nucleon medium,  but when the temperature is closer to the critical temperature, the effective nucleon mass decreases very sharply and this also accompanies with the sudden increasing of the radium of the nucleon. Eventually, at some critical temperature we find no more a localized soliton solution and nucleons lose their individual character, which signals a deconfinement phase transition from the nucleon to quark matter. It is worth to point out that the results in the present work are qualitatively in agreement with the work of Bernard and Meissner\cite{Bernard:1989pj}, in which they have investigated the static and dynamic properties of the nucleon by using a chiral soliton model with explicit vector mesons at finite temperature. 

Although the nucleons' static and dynamics properties at finite temperature and baryon chemical potential show some similar behaviors both in a hot nucleon medium and in a hot quark medium, the phase structures are quite different according to these two different scenarios. In the case of a quark medium, there is a crossover in the low baryon chemical potential and a first-order phase transition in the high baryon chemical potential. However, in the case of nucleon medium, there always exhibits a crossover in the whole $T-\mu_B$ phase plane, and in contrast to the case of a quark medium where the phase boundary signaled by the dramatic structure change of the nucleon, the deconfinement phase boundary in the case of nucleon medium is characterized by the moment when nucleons start to overlap.  Moreover, in comparison with the QCD phase structure of the PQM model in a quark medium shown as Ref.~\cite{Jin:2015goa}, when the baryon chemical potential is less than $220$ MeV, the dramatic structure change of the isolated nucleon and the overlap of nucleons are going to happen almost simultaneously, but when the baryon chemical potential is larger than $220$ MeV, the soliton model based on the PQM model predicts that nucleons start to overlap with each other near the deconfinement phase boundary, and valence quarks inside the single nucleon are partly deconfined due to the fact that nucleons now can share in valence quarks with others. How to search, study and identify this kind of ``leaking'' quark phase would be interested in the relativistic heavy Ion collision experiments in future.

There are many avenues for further investigation: the bulk thermodynamical description of the hadron-quark phase transitions when considering the effect of the overlap of the nucleon, including vector and axial-vector mesons, such as $\rho$, $\omega$, and $a$ in the PQM model~\cite{Broniowski:1985kj}, nucleon properties based on the nontopological model at finite nuclei in hadron phase, and so on.

Especially, the most interesting study is to include the vector and axial-vector mesons in the PQM model, because such extension can low down the soliton energy to a reasonable and acceptable range about $939$ MeV~\cite{Alkofer:1990ih,Weigel:1994sz,Weigel:2008zz}, which is often taken as a standard parameter in the RMF model. In other words, in the present form of study, we still face a self-consistent problem of that the mass of a free nucleon as a soliton is relatively larger than the TM1 parameter set in our calculations. We believe that the work in this direction will cure the problem and until then the equation of state can be consistently applied to describe the hadron-quark phase transition in a dense matter.

\begin{acknowledgments}
We thank Jinshuang Jin and Xiongjie Wang for valuable comments and discussions. This work was supported in part by National Natural Science Foundation of China (NSFC) under No.11675048 and No. 11775119.
\end{acknowledgments}

\end{document}